\documentclass[pre,aps,twocolumn,showpacs,amsmath,amssymb,amsfonts,floatfix,superscriptaddress]{revtex4}
\usepackage{graphicx}
\usepackage{dcolumn}
\usepackage{bm}
\usepackage{latexsym}
\usepackage{float}
\usepackage{supertabular}
\usepackage{longtable}

\begin{document}
\title{Evolutionary games with facilitators: When does selection favor cooperation?}
\author{Mauro Mobilia}
\affiliation{Department of Applied Mathematics, School of Mathematics, University of Leeds, Leeds LS2 9JT, U.K.} %
\email{M.Mobilia@leeds.ac.uk}
\begin{abstract}
We study the combined influence of selection and random fluctuations on the evolutionary dynamics of 
two-strategy (``cooperation'' and ``defection'') games in populations comprising cooperation facilitators. The latter are individuals that
support cooperation by enhancing the reproductive potential of cooperators relative to the fitness of defectors.
By computing the fixation probability of a single cooperator in 
 finite and well-mixed populations that include a fixed number of facilitators,
and by using mean field analysis,
we determine when selection promotes cooperation   in the important
 classes of prisoner's dilemma, snowdrift and stag-hunt games.
 In particular, we identify the circumstances under which selection favors the replacement and invasion of
defection by  cooperation.
Our findings, corroborated by stochastic simulations, show that the spread of 
cooperation can be promoted through various scenarios when the density of facilitators exceeds a critical value whose dependence 
on the population size and selection strength is analyzed.
We also determine under which conditions cooperation is more likely to replace defection than vice versa.

\end{abstract}

\maketitle

\section*{1. Introduction}
%
Evolutionary game theory (EGT)~\cite{EGT1,EGT2,EGT3,EGT4,EGT5,EGT6} provides  a suitable framework to model
the evolution of cooperative behavior, which is a  problem of paramount importance~\cite{Pennisi}. EGT is traditionally concerned with 
the dynamics of competing species (strategies) in infinitely large populations
where each species' reproductive potential (fitness) varies with the population composition.
The selection pressure is thus ``frequency-dependent'' and
the deterministic evolution is  described in terms of ordinary differential (replicator) equations~\cite{EGT1,EGT2,EGT3,EGT4,EGT5}. 
The classic notion of evolutionary stable strategy (ESS) 
is related to the stability of the fixed points of the underlying replicator equations: a strategy is evolutionary
stable  if the fitness of those adopting it 
is always higher than the fitness of a small fraction of mutants with an alternative strategy.
As an important consequence of frequency-dependent selection, the fitness optimization at an individual level
may cause the reduction of the population's average fitness. This yields a ``social dilemma'', as in the 
celebrated prisoner's dilemma game where the best strategy for each player is to defect even if mutual cooperation
would benefit the whole population~\cite{PD,EGT4,EGT5,EGT6}. Cooperation  dilemmas can arise in other simple models, like
the two-strategy snowdrift and stag-hunt games~\cite{EGT4,EGT5,EGT6}. The attempt to explain the cooperative behaviors observed 
in experiments~\cite{PDexp1,PDexp2}
has motivated the investigation of various mechanisms (kin and group selection, reciprocity, local interactions)
that could possibly promote the spread of cooperation~\cite{v1,v2,v3,v4,v5,v6,v7,v8,v9,v10,v11,v12}.

When the size of the population is finite, evolution is known to be affected by demographic noise. 
The dynamics is thus often modeled in terms of continuous-time Markov chains
and characterized by the notion of fixation probability~\cite{EGT4,EGT5,EGT6}. 
This concept refers to the possibility that a ``mutant type''
takes over the entire population and is closely related to the important concept
of evolutionary stability in finite populations (ESS$_N$) put forward by Nowak and collaborators~\cite{ESSN}.
These authors proposed that in two-player games a given strategy (say $\textsf{D}$) is
 evolutionary stable in a finite population if, in addition to having a higher 
fitness than the alternative strategy (say $\textsf{C}$), the selection also opposes the replacement
of $\textsf{D}$ by $\textsf{C}$. 
As a direct consequence, when the fixation probability of the mutant type $\textsf{C}$
is higher than in the absence of selection, the evolution favors $\textsf{C}$ replacing  $\textsf{D}$~\cite{EGT4,EGT5,EGT6,ESSN}.

In this work, we consider the evolutionary dynamics of three important classes of two-strategy games, namely
the prisoner's dilemma (PD), snowdrift (SD) and stag-hunt (SH) games, where the pure strategies are
``cooperation'' ($\textsf{C}$) 
and ``defection'' ($\textsf{D}$) and the former is supported by the presence  of a number of
``cooperation facilitators'' in the population. 
The facilitators participate in the dynamics by enhancing
the fitness of cooperators relative to the reproductive potential of defectors~\cite{MM12}.
Here, our goal is to analyze the interplay in finite populations between selection, demographic fluctuations and facilitators
in these paradigmatic classes of evolutionary games. 
In fact, it is important to understand whether selection favors cooperation, and in particular when it promotes 
the replacement of defectors by cooperators~\cite{EGT1,EGT2,EGT4,EGT5,EGT6,ESSN}. For instance, while selection always opposes cooperation in the classic
PD games (no facilitators), it has been found that cooperation can spread in a special class of PD games with 
facilitators~\cite{MM12}. 
For  PD, SD and SH games,
 we here investigate the circumstances and scenarios under which selection and the presence of facilitators favor cooperation at 
 the expense of defection, and how these conditions are influenced by the population size and 
selection intensity. To understand  when the evolutionary dynamics promotes the spread of cooperation, we determine  the critical fraction of 
facilitators in the population above which selection favors the replacement of defection by cooperation, as well as the 
density of facilitators necessary for selection to favor cooperation invading defection. 
We also  identify the  conditions under which cooperation is more likely to replace defection than vice versa.

The structure of this paper is the following: In the next section, we define the prisoner's dilemma, snowdrift and stag-hunt 
games in the presence of cooperation facilitators. Section 3 is dedicated to the study of the dynamics in finite 
populations and to the calculation of the fixation probability. 
The circumstances under which selection favors the invasion and the 
replacement of defection by cooperation are studied in Section 4-6. Our conclusions are presented in Section 7.

\section*{2. Evolutionary games with cooperation facilitators}
We study the evolutionary dynamics of two-strategy games, that can be interpreted as social dilemmas, in the presence of 
 cooperation facilitators. More specifically, we consider  a {\it finite} and well-mixed (spatially homogeneous) population 
consisting of $N$ individuals, $j$ of which are cooperators ($\textsf{C}$'s) and $k$ are 
defectors ($\textsf{D}$'s). The population is also comprised of a fixed  number $\ell$ 
of ``cooperation facilitators'' that enhance the expected payoff of $\textsf{C}-$players 
relative to the  average of defectors (see below)~\cite{MM12}.
The  pairwise interaction between $\textsf{C}$'s and $\textsf{D}$'s is embodied in the 
payoff matrix 
and we distinguish three classes of games: the (i) {\it prisoner's dilemma} (ii)
{\it snowdrift}, (iii) and {\it stag-hunt} games~\cite{EGT1,EGT2,EGT3,EGT4,EGT5,EGT6,PD,snowdrift1,snowdrift2,snowdrift3,staghunt1,staghunt2}.

(i) The paradigmatic model of cooperation dilemma is the  {\it prisoner's dilemma} (PD) game
whose generic features are conveniently captured by the  following payoff 
matrix~\cite{EGT1,EGT2,EGT3,EGT4,EGT5,EGT6,PD} with three parameters:
\begin{eqnarray}
 \label{PM_PD}
\bordermatrix{
  & \textsf{C} & \textsf{D} \cr
\textsf{C} & b-c & -c \cr
\textsf{D} & b & d-c \cr},
\end{eqnarray} 
where $b>c>0$ and $b>d>0$. The parameters $b$ and $c$ respectively represent the benefit and the cost of cooperation
and, in the context of PD games, $r\equiv c/b$ is referred to as the ``cost-to-benefit ratio''~\cite{MM12}. 
Hence, when  one player defects and the other cooperates, the latter is ``exploited'' and receives a payoff $-c$  
while the former gets $b$. The payoff
for mutual  defection is given by $d-c$ and can be either non-negative ($c\leq d$) or negative ($c> d$), but always 
less than the payoff ($b-c$) for mutual cooperation and superior to the exploitation payoff ($-c$). 
To simplify our analysis, we shall always consider that $b,c,d$ are all of order $O(1)$.
 Defection is the only strict Nash equilibrium (NE) 
of the classic ($\ell=0$) PD games and a cooperation dilemma arises from 
the fact that each individual favors defection, even  though cooperation enhances the overall payoff of the population.
In the presence of a fixed number $\ell=N-j-k$ of cooperation facilitators, the expected payoffs 
of cooperators and defectors, respectively $\pi_{\textsf{C}/\textsf{D}}$, are inferred from (\ref{PM_PD}) according to the general 
principles of EGT, yielding (by excluding self-interactions)~\cite{MM12}:
\begin{eqnarray}
 \label{payoffs_PD}
\pi_{\textsf{C}}&=& (b-c)\left(\frac{j+\ell-1}{N-1}\right)-c \left(\frac{N-\ell-j}{N-1}\right)\nonumber\\
\pi_{\textsf{D}}&=& b \left(\frac{j}{N-1}\right)+(d-c) \left(\frac{N-\ell-j-1}{N-1}\right).
\end{eqnarray} 
As facilitators  enhance $\pi_{\textsf{C}}$ by $b\ell/(N-1)$ and
$\pi_{\textsf{D}}$ by $(c-d)\ell/(N-1)$, their presence supports cooperation relative to defection (since $b>c-d$), 
see also (\ref{deltapi})-(\ref{beta}).
The population average payoff is  given by $\bar{\pi}=(j\pi_{\textsf{C}}+
k\pi_{\textsf{D}})/N$.  
The fixation properties of the special subclass of PD games with facilitators and $c=d$ was 
 studied in Ref.~\cite{MM12}.

(ii)  {\it Snowdrift} (SD) games are the prototypes of evolutionary models exhibiting mixed evolutionary stable strategy.
The  features of these anti-coordination games  are aptly captured by the following payoff 
matrix~\cite{EGT1,EGT2,EGT3,EGT4,EGT5,EGT6,snowdrift1,snowdrift2,snowdrift3}:
\begin{eqnarray}
 \label{PM_SD}
\bordermatrix{
  & \textsf{C} & \textsf{D} \cr
\textsf{C} & b-c & d-c \cr
\textsf{D} & b & -c \cr},
\end{eqnarray} 
where we  consider that  $b>c+d>0$, with $c>0$ and $d>0$. In its classic version ($\ell=0$), SD games are characterized by an evolutionary 
stable (non-strict NE) mixed strategy where ${\textsf{C}}$ is played with frequency $c/(c+d)$.

When a  fixed number $\ell$ of cooperation facilitators is present in the population, the expected payoffs 
of cooperators and defectors are inferred from (\ref{PM_SD}): 
\begin{eqnarray}
 \label{payoffs_SD}
\pi_{\textsf{C}}&=& (b-c)\left(\frac{j+\ell-1}{N-1}\right) +(d-c) \left(\frac{N-\ell-j}{N-1}\right)\nonumber\\
\pi_{\textsf{D}}&=& b \left(\frac{j}{N-1}\right) -c \left(\frac{N-\ell-j-1}{N-1}\right),
\end{eqnarray} 
where the presence of facilitators enhances $\pi_{\textsf{C}}$ by $(b-d)\ell/(N-1)$ and
$\pi_{\textsf{D}}$ by $c\ell/(N-1)$, and therefore promotes cooperation relative to defection (since $b-d>c$).

(iii) {\it Stag-hunt} (SH) games are the prototype of social contrast and trust games. 
Many  features of these coordination games are suitably captured by the simple payoff 
matrix~\cite{EGT1,EGT2,EGT3,EGT4,EGT5,EGT6,staghunt1,staghunt2,weaksel1,weaksel2}:
\begin{eqnarray}
 \label{PM_SH}
\bordermatrix{
  & \textsf{C} & \textsf{D} \cr
\textsf{C} & b & -c \cr
\textsf{D} & b-c & d-c \cr},
\end{eqnarray} 
where $b>c>d>0$. In the absence of facilitators, SH games are characterized by two strict pure
NEs and by a non-strict NE 
corresponding to  a mixed strategy in which ${\textsf{C}}$ is played with frequency $d/(c+d)$.

In the presence of  $\ell$  cooperation facilitators, the expected payoff 
of cooperators and defectors
are obtained from (\ref{PM_SH}):
\begin{eqnarray}
 \label{payoffs_SH}
\pi_{\textsf{C}}&=& b\left(\frac{j+\ell-1}{N-1}\right) -c \left(\frac{N-\ell-j}{N-1}\right)\\
\pi_{\textsf{D}}&=& (b-c) \left(\frac{j}{N-1}\right)+(d-c) \left(\frac{N-\ell-j-1}{N-1}\right),\nonumber
\end{eqnarray} 
where the presence of facilitators enhances $\pi_{\textsf{C}}$ by $(b+c)\ell/(N-1)$ and
$\pi_{\textsf{D}}$ by $(c-d)\ell/(N-1)$, and therefore supports cooperation relative to cooperation.

For further convenience, it is useful to write the difference $\delta \pi(j)$
between the expected payoffs of defectors and cooperators for the three types of games (PD, SD and SH)  as
\begin{eqnarray}
 \label{deltapi}
\delta \pi(j)=\pi_{\textsf{D}}(j)-\pi_{\textsf{C}}(j)=\alpha\left(j/N\right)+\beta_{\ell},
\end{eqnarray} 
with
\begin{eqnarray}
 \label{alpha}
 \alpha=\left\{
  \begin{array}{l l l}
   (c-d)\frac{N}{N-1}  & \quad \text{(PD)}\\
    (c+d)\frac{N}{N-1} & \quad \text{(SD)}\\
     -(c+d)\frac{N}{N-1} & \quad \text{(SH)}\\
  \end{array} \right.
\end{eqnarray}
and
\begin{eqnarray}
 \label{beta}
 \beta_{\ell}=\left\{
  \begin{array}{l l l}
   \frac{dN-(b+d-c)\ell+(b-d)}{N-1}  & \quad \text{(PD)}\\
    \frac{b-dN-(b-c-d)\ell}{N-1} & \quad \text{(SD)},\\
     \frac{dN-(b+d)\ell+b+c-d}{N-1} & \quad \text{(SH)}\\
  \end{array} \right.
\end{eqnarray}
where we have used (\ref{payoffs_PD}), (\ref{payoffs_SD}), (\ref{payoffs_SH}).
Since facilitators enhance $\pi_{\textsf{C}}$ relative to $\pi_{\textsf{D}}$, we
 notice that for all the  above games  $\delta \pi(j)$ and $\beta_{\ell}$ are decreasing functions of $\ell$.

\section*{3. Dynamics, cooperation fixation probability and evolutionary stability}
In a population of finite size $N$, the evolutionary dynamics is generally described 
by a continuous-time birth-death process~\cite{EGT4,EGT6,Gardiner,vanKampen,weaksel3}, where 
at each time step a pair of individuals is randomly drawn to interact and the 
competition between the strategies, specified by the underlying payoff matrix, is encoded
in the transition rates: if one picks a cooperator-defector 
pair, one of these individuals
is randomly chosen for reproduction proportionally to its expected payoff (or fitness) 
and the other is replaced by the newborn offspring. In each elemental move, the population composition
evolves from a state with $j$ cooperators to a state consisting of $j\pm 1$ individuals of type $\textsf{C}$.
The underlying Markov chain is fully specified by the transition rates 
$T_j^{\pm}$ for the moves $j \to j\pm 1$, respectively, with $T_{j= N-\ell}^{\pm}=T_{j=0}^{\pm}=0$ to account
for the absorbing states $j=0$ (no cooperators) and $j=N-\ell$ (no defectors)~\cite{MM12}.
More specifically, we  here consider the transition rates~\cite{EGT6,FermiProcess1,FermiProcess2,FermiProcess3,FermiProcess4,snowdrift2,MM12}:
\begin{eqnarray}
 \label{rates}
T_j^{\pm}=\frac{j(N-\ell-j)}{N(N-1)}\, \frac{1}{1+e^{\pm\{f_{\textsf{D}}(j) - f_{\textsf{C}}(j)\}}}.
\end{eqnarray}
In these expressions $j(N-\ell-j)/N(N-1)$ accounts for 
the 
probability of picking a cooperator-defector pair. The probability that a $\textsf{C}$ reproduces and replaces a  $\textsf{D}$ is proportional
to $[1+e^{\{f_{\textsf{D}}(j) - f_{\textsf{C}}(j)\}}]^{-1}$ (and similarly for $\textsf{D}$ reproducing at the expense 
of $\textsf{C}$). This corresponds to the evolution according to the so-called
comparison pairwise ``Fermi process''
(FP)~\cite{EGT5,EGT6,FermiProcess1,FermiProcess2,FermiProcess3,FermiProcess4,snowdrift2,snowdrift3,staghunt1,staghunt2,weaksel1,weaksel2}, where 
$f_{\textsf{C}}$ and $f_{\textsf{D}}$ respectively denote the cooperators and defectors fitness. These quantities 
are commonly
chosen  to be proportional to the difference between the $\textsf{C/D}$'s expected payoffs and the 
average population payoff~\cite{EGT6,MM12,staghunt1,staghunt2,weaksel1,snowdrift2,snowdrift3}:
\begin{eqnarray}
 \label{fitness}
f_{\textsf{C/D}}(j)=1-s+s[\pi_{\textsf{C/D}}(j)-\bar{\pi}(j)],
\end{eqnarray}
where the parameter $0\leq s\leq 1$ accounts for the selection strength ($s=0$ and $s=1$ respectively 
correspond to the weak and strong selection  limits) and the $1-s$ is a baseline fitness contribution~\cite{EGT6,staghunt1,staghunt2,weaksel1,weaksel2,snowdrift2}. 
It has to be noted that $f_{\textsf{C/D}}(j)$ vary with the population composition 
(``frequency-dependent selection''~\cite{EGT1,EGT2,EGT3,EGT4,EGT5,EGT6}) and, as before, it is useful to write the 
difference between 
$f_{\textsf{D}}$ and $f_{\textsf{C}}$ as $\delta f(j)=f_{\textsf{D}}-f_{\textsf{C}}=(s/N)[\alpha j +N\beta_{\ell}]$.
It is also worth mentioning that 
other microscopic update rules, like the Moran process (MP)~\cite{Moran,EGT4,EGT5,EGT6,Antal06,snowdrift1,snowdrift2,weaksel1,weaksel2,weaksel3}, 
can also  be chosen. The choice of the FP is made  to simplify the analysis [(\ref{rates}) only depends on 
$\delta f(j)$], but it is noteworthy 
that the FP and MP are known to generally lead to qualitatively similar dynamics and, in the biological relevant 
case of weak selection intensity ($s\ll 1$), they even predict quantitatively similar results~\cite{EGT6,weaksel1,weaksel2,weaksel3,snowdrift2,MM12}.

In the limit of an infinitely large population ($N\to\infty)$, with 
$x\equiv j/N$,  $z\equiv \ell/N$ and $T_{j}^{\pm}\to T^{\pm}(x)$ and $\beta_{\ell}\to \beta(z)$, the random fluctuations are negligible and the dynamics is aptly described
by the mean field rate equation (RE, after proper rescaling of time)
\begin{eqnarray}
 \label{MF}
 (d/dt)x(t)&\equiv& \dot{x}= T^+(x) -  T^-(x) \\
 &=&-x(1-z-x)\tanh{\left\{\frac{s}{2}(\alpha x +\beta(z))\right\}}.\nonumber
\end{eqnarray}
This replicator-like equation is characterized by 
two absorbing fixed points, $x=0$ (no $\textsf{C}$'s) and  $x=1-z$ (no $\textsf{D}$'s), and 
possibly by an interior fixed point $x^*=-\beta(z)/\alpha$ corresponding to 
the coexistence of $\textsf{C}$'s and $\textsf{D}$'s (when $0<x^*<1$).

When the population size is finite ($N<\infty$) and random fluctuations cannot be disregarded, the 
evolutionary dynamics is often characterized by the cooperation fixation probability $\phi^{\textsf{C}}_j$ which,
in the presence of $\ell$ cooperation facilitators, is the probability that an initial number of $j$ cooperators eventually
replace all the defectors, leading to the absorbing state consisting of $N-\ell$ cooperators and $\ell$ facilitators.
This quantity satisfies the  (backward) master equation, $j\in [0, N-\ell]$:
\begin{eqnarray}
 \label{backME}
T_j^-\phi_{j-1}^{\textsf{C}}+T_j^+\phi_{j+1}^{\textsf{C}}-[T_j^- + T_j^+]\phi_j^{\textsf{C}}=0, 
\end{eqnarray} 
with $\phi_0^{\textsf{C}}=0$ and $\phi_{N-\ell}^{\textsf{C}}=1$. 
With (\ref{rates}), the solution of (\ref{backME})
reads (see, e.g.,~\cite{vanKampen,MM12,EGT4,EGT6}):
\begin{eqnarray}
 \label{Sol}
\phi_j^{\textsf{C}}=\frac{1+\sum_{n=1}^{j-1}  {\rm exp}\left(\frac{sn}{2N}[\alpha(n+1)+2N\beta_{\ell}] \right)}
{1+\sum_{n=1}^{N-\ell-1}{\rm exp}\left(\frac{sn}{2N}[\alpha(n+1)+2N\beta_{\ell}] \right)}.
\end{eqnarray} 
In particular, the fixation probability of a single cooperator in a sea of $N-\ell -1$ defectors and in the presence
of $\ell$ cooperation facilitators is $\phi_1^{\textsf{C}}\equiv \phi^{\textsf{C}}$, with
\begin{eqnarray}
 \label{phi1C}
\phi^{\textsf{C}}=\left[
1+\sum_{n=1}^{N-\ell-1}{\rm exp}\left(\frac{sn}{2N}[\alpha(n+1)+2N\beta_{\ell}] \right)\right]^{-1}.
\end{eqnarray} 
Clearly, $\phi^{\textsf{C}}=1$ when $\ell=N-1$. It has also to be noted that in the absence of selection ($s=0$),
the fixation probability of $j$ cooperators  for the ensuing neutral dynamics simply 
reads $\phi_j^{\textsf{C}}=j/(N-\ell)$, where $N-\ell$ can be interpreted as a population size effectively
reduced by the presence of $\ell$ facilitators.

The notion of fixation probability is particularly important because it is closely related to the
concept of evolutionary stability in finite populations~\cite{ESSN,EGT4,EGT5,EGT6}, and it allows to study under which circumstances the spread 
of cooperation is favored by selection, as proposed by the authors of Ref.~\cite{ESSN}. 
This line of reasoning is readily generalized in the presence of cooperation facilitators: 
Defection is an evolutionary stable strategy  in a population of size $N$ 
comprising  $\ell$ cooperation facilitators (ESS$_{N}^{\ell}$)
 if the following {\it two conditions} are {\it simultaneously} satisfied:
\begin{enumerate}
 \item[i.] A single (``mutant'') cooperator must have a
lower fitness than (``resident'') defectors,  i.e. $\delta f(1)=(s/N)[\alpha + N\beta_{\ell}]>0$.
 \item[ii.] For $0\leq \ell<N-1$, the fixation probability of a single cooperator has to be  
  lower than in the absence of selection ($s=0$, when there are only random fluctuations), i.e. 
  $\phi^{\textsf{C}}<(N-\ell)^{-1}$. 
\end{enumerate}
The condition (i) ensures that selection opposes the initial spread of cooperation.
Furthermore, as the fixation of a strategy can be favored in finite populations 
even if its initial increase is opposed, the 
condition (ii) guarantees that selection opposes the replacement of defection by cooperation.

According to Ref.~\cite{ESSN} (see also \cite{EGT4,EGT6}), when any of the above conditions (i), (ii) is violated, defection is not an ESS$_{N}^{\ell}$. 
If the  condition (i) is satisfied but $\phi^{\textsf{C}}>(N-\ell)^{-1}$, 
selection  favors the replacement but opposes the
invasion of defection by cooperators: in this case, the fixation of $\textsf{C}$ is
favored  even if the initial increase of $\textsf{C}$'s frequency is opposed.  This typically occurs in SH games (see Sec.~6). 
On the other hand, when the  condition (ii) is fulfilled but  $\delta f(1)<0$, selection  
favors the invasion by cooperators but opposes the replacement 
of defection by cooperation. 

As we are here chiefly interested in determining when selection promotes cooperation in the presence
of facilitators, we need to understand when the conditions favoring the invasion and replacement by cooperation are 
satisfied.
With the definitions (\ref{alpha},\ref{beta}) of $\alpha$
and $\beta_{\ell}$, together with (\ref{fitness}) and (\ref{Sol}), the {\it invasion condition} $\delta f(1)<0$
for which selection favors the invasion of defectors by cooperators  reads
\begin{eqnarray}
\label{invasion}
 \alpha+N\beta_{\ell}<0,
\end{eqnarray}
while the {\it replacement condition} $\phi^{\textsf{C}}>(N-\ell)^{-1}$ under which selection favors the replacement of defection by cooperation
(for $0\leq \ell<N-1$) is
\begin{eqnarray}
\label{replacement}
N-\ell-1&>&\sum_{n=1}^{N-\ell-1}e^{n \psi_{\ell}(n)}, \quad \text{with} \\
\label{replacement-2}
 \psi_{\ell}(n)&=&\delta f(1)+s\alpha \left(\frac{n-1}{2N}\right). \nonumber
\end{eqnarray}
In the next sections we analyze the implications of the  conditions (\ref{invasion}) and (\ref{replacement}) 
in PD, SD and SH games. However,
 since  (\ref{invasion}) imposes $\delta f(1)<0$, we can already notice that 
 (\ref{replacement}) is automatically satisfied when 
 the invasion condition $\beta_{\ell}<-\alpha/N$ is fulfilled and $\alpha\leq 0$.

 Furthermore, in order to understand  whether the strategy ${\textsf{C}}$ is more likely
to replace ${\textsf{D}}$ than vice versa, it may be  interesting to compare $\phi^{\textsf{C}}$ with the probability
 $\phi^{\textsf{D}}$  that a single defector  $\textsf{D}$  takes over a population 
 consisting of  $N-\ell-1$ cooperators $\textsf{C}$ and $\ell$ cooperation facilitators~\cite{EGT4,EGT6}.
 Since fixation in either state $j=N-\ell$ or $j=0$ is guaranteed, one has  
 $\phi^{\textsf{D}}=1-\phi_{N-\ell-1}^{\textsf{C}}$ and, with (\ref{Sol}), one readily  obtains
 \begin{eqnarray}
  \label{phiCphiD}
  \frac{\phi^{\textsf{D}}}{\phi^{\textsf{C}}}&=&
  {\rm exp}\left(s\sum_{j=1}^{N-\ell-1}
  \delta f(j)
  \right)\nonumber\\
  &=& {\rm exp}\left(\frac{s(N-\ell-1)}{2N}[\alpha(N-\ell)+2N\beta_{\ell}] \right).
 \end{eqnarray}
As a result, $\phi^{\textsf{C}}>\phi^{\textsf{D}}$ when $s>0$ and
$\alpha(1-z)+2\beta_{\ell}<0$. In large populations ($N\gg 1$), $\beta_{\ell}\to \beta(z)=-\alpha x^*$
and one finds $\phi^{\textsf{C}}>\phi^{\textsf{D}}$ if $x^*<(1-z)/2$ when $\alpha<0$, which corresponds to 
$\textsf{C}$ having a larger basin of attraction than $\textsf{D}$ and being risk-dominant~\cite{EGT4,EGT6}.

\section*{4. When does selection favor cooperation in prisoner's dilemma games with facilitators?}
In PD games with $\ell$ cooperator facilitators the expected payoffs of cooperators and 
defectors  are given by (\ref{payoffs_PD}), which yields the
fitness difference 
$\delta f(j)=s[\alpha (j/N) + \beta_{\ell}]$
 with  $\alpha=(c-d)N/(N-1)$ and $\beta_{\ell}=[dN-(b+d-c)\ell+(b-d)]/(N-1)$. 

It is instructive to first consider the deterministic mean field limit ($N\to \infty$, with $c\neq d$)
 in which the dynamics is described by the RE (\ref{MF}). The latter is characterized
by the system's two absorbing  fixed points $x=0$ (no $\textsf{C}$'s) and $x=1-z$ (no $\textsf{D}$'s).
When $0<-\beta(z)/\alpha<1$, these are separated by an interior fixed point $x^*=[(b+d-c)z-d]/(c-d)$. A 
stability analysis
of (\ref{MF}) reveals various mean field  scenarios:
\begin{itemize}
\item If $c>d$ and $z\leq d/(b-c+d)$, there is no interior fixed point and $x=0$ is the only attractor of the RE (\ref{MF}). 
This corresponds to the situation where there are no cooperators in the final state.
 \item If $c>d$ and $d/(b-c+d)<z\leq r$, where  $r=c/b$ is the 
cost-to-benefit ratio, $x^*=[(b+d-c)z-d]/(c-d)$
is the only attractor of the RE (\ref{MF}). In this case, 
the presence of facilitators leads to the coexistence of 
cooperators and defectors.
\item If $c>d$ and $z>r$,  $x=1-z$ is the only attractor of the RE (\ref{MF}).
This corresponds to the situation where (\ref{MF}) predicts that defectors are replaced by cooperators.
\item If $c<d$ and $z\leq r$,  $x=0$ is the only attractor of the RE (\ref{MF}).
\item If $c<d$ and $r<z\leq d/(b-c+d)$, both $x=0$ and $x=1-z$ are stable and separated by the unstable interior fixed point 
$x^*$. This situation is characterized by bistability and, depending on whether the initial density of  $\textsf{C}$'s is 
more or less than $x^*$, the final state will either be  $x=1-z$ or $x=0$~\cite{EGT2,EGT3,EGT4,EGT6}. 
\item If $c<d$ and $z>d/(b-c+d)$, $x=1-z$  is the only attractor of the RE (\ref{MF}) and 
this corresponds to the eventual replacement of defectors by cooperators.
\end{itemize}
It has to be noted that the special case $c=d$ corresponds to the ``equals-gains-from-switching'' PD game~\cite{EGT6}
discussed in Ref.~\cite{MM12}.

It appears from the above discussion that the existence and stability of
$x^*$, as well as the sign of $c-d$ (payoff for mutual defection), drastically influence 
the spread of cooperation. In fact, when $z\leq d/(b-c+d)$ there is no interior fixed point in the case $c>d$ whereas
$x^*$ exists but is unstable when $c<d$ and $z>r$. According to this mean field picture, we expect that  a small number of
cooperators will quickly die out if $z\leq d/(b-c+d)$. It is therefore necessary that $z>d/(b-c+d)$ for cooperation 
to be able to prevail when $N\to \infty$. In fact,  when $z>d/(b-c+d)$ and $c<d$  we certainly expect that
selection favors cooperation since  $x=1-z$ is then the attractor of  (\ref{MF}). 
The situations where $d/(b-c+d)<z\leq r$ with $c>d$,
and $z\leq r$ when $c< d$,  are more intriguing because 
(\ref{MF}) is thus  characterized by a stable coexistence fixed point $x^*$ and it is not obvious
under which circumstances cooperation will be able to replace defection. In fact, below we show that the 
replacement of ${\textsf{D}}$ by ${\textsf{C}}$
is favored when $z>z^*\equiv \ell^*/N$, where $d/(b-c+d)\leq z^*\leq r$ when $c>d$ 
and  $r \leq z^*\leq d/(b-c+d)$ when $c\leq d$.

 Building on the insight gained from the above mean field analysis, we now assess the combined influence of 
 the selection pressure and demographic fluctuations in finite populations, by  
 considering  the invasion condition (\ref{invasion}) that reads $z>\widetilde{z}$ with
 \begin{eqnarray}
  \label{invasionPD}
  \widetilde{z}\equiv \widetilde{\ell}/N=\frac{d}{b+d-c}+\left(\frac{b+c-2d}{b+d-c}\right)\frac{1}{N},
 \end{eqnarray}
as well the replacement condition (\ref{replacement}) which yields
 \begin{eqnarray}
  \label{replacementPD}
  N-\ell-1&>&\sum_{n=1}^{N-\ell-1}e^{n \psi_{\ell}(n)}, \quad \text{with}\\
  \label{replacementPD-2}
  \psi_{\ell}(n) &=&\delta f(1)+s\frac{(c-d)(n-1)}{2(N-1)}\nonumber\\
  &=&s \left[\frac{(c-d)(n+1)}{2(N-1)}+\beta_{\ell}\right].
 \end{eqnarray}
We notice that in the limit $N\gg 1$ the condition $z>\widetilde{z}$ is equivalent to $z>d/(b+d-c)$.

With (\ref{phiCphiD}), (\ref{alpha}) and (\ref{beta}), 
one finds that cooperation is more likely to replace defection than vice versa when   
\begin{eqnarray}
\label{phiCphiDPD}
z>\frac{c+d}{2b+d-c}+\frac{2}{N}\left(\frac{b-c}{2b-c+d}\right).
\end{eqnarray}

To analyze when selection favors cooperation in PD games with $\ell$ cooperators whose 
expected payoffs is given by (\ref{payoffs_PD}),
 one needs to distinguish the cases where the payoff 
for mutual defection is negative ($c>d$)
and non-negative. 
 \subsection*{4.1. The case of negative payoff for mutual defection $(c>d)$}
 When there is a negative payoff $d-c<0$ for mutual defection, 
 $\alpha>0$
 and the replacement condition (\ref{replacementPD}) is generally more stringent
 than $z>{\widetilde z}$ with (\ref{invasionPD}).

 In fact, there is generally  an $\ell^*$,
 with  $\lfloor \widetilde{\ell}\rfloor\leq \ell^* \leq \lceil Nr \rceil$, such that 
$\sigma_{\rm PD}(N,\ell)=
\sum_{n=1}^{N-\ell-1}e^{n \psi_{\ell}(n)}-(N-\ell-1)< 0$
   when $ \ell > \ell^*$ and $s>0$~\cite{ValInt}.
\begin{figure}
\includegraphics[width=3.6in, height=2.4in,clip=]{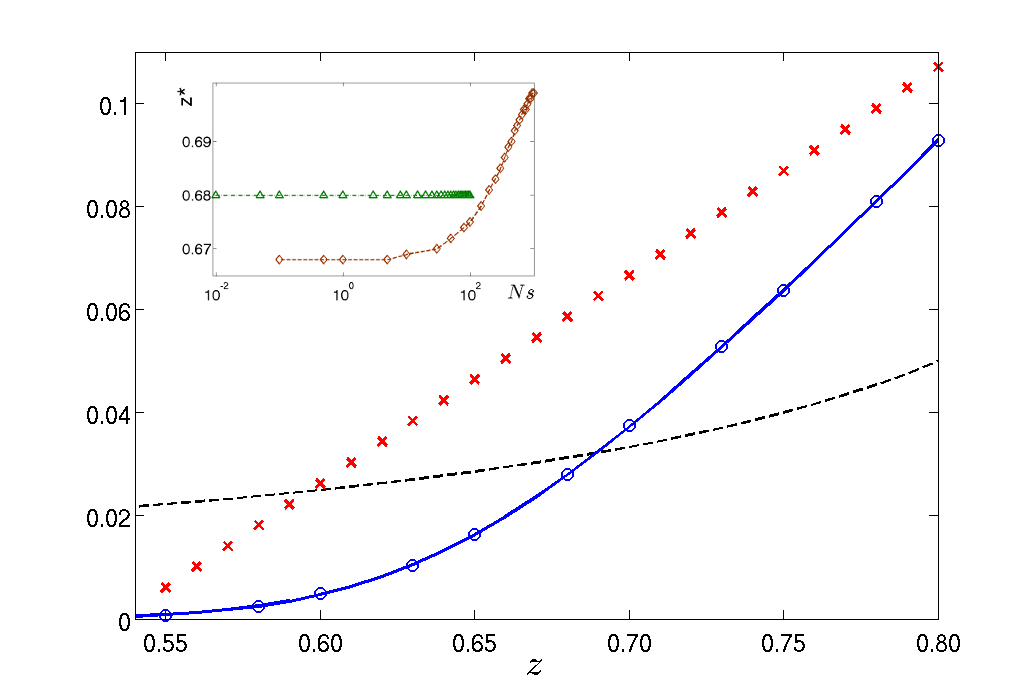}
\includegraphics[width=3.6in, height=2.4in,clip=]{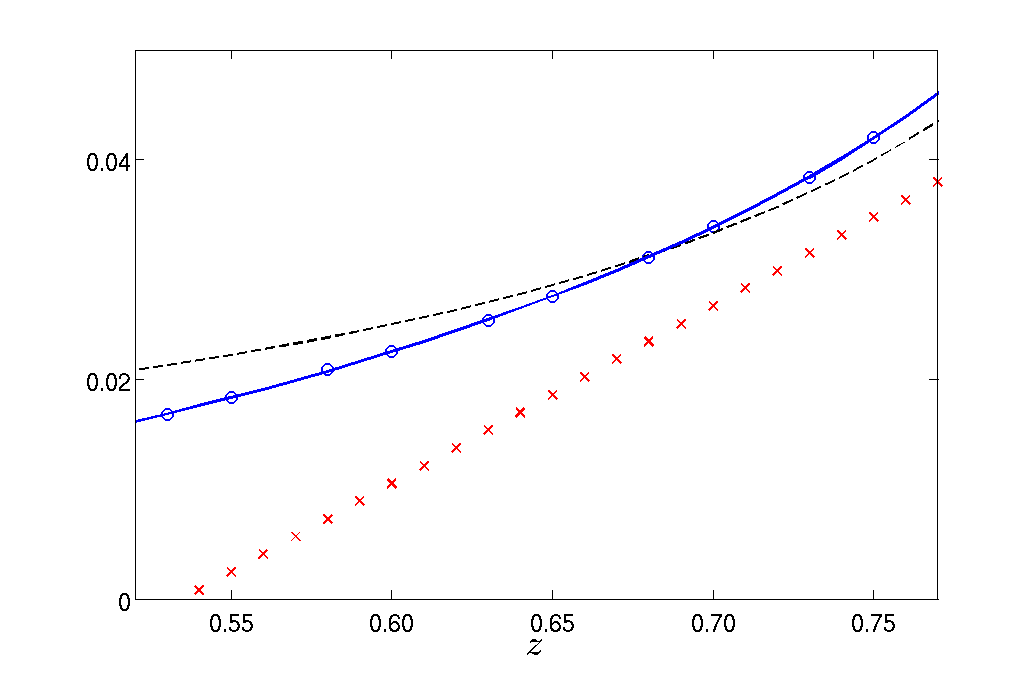}
\caption{{\it (Color online)}. 
$\phi^{\textsf{C}}$ and $-\delta f(1)>0$ vs. $z=\ell/N$
in PD games with  
$\ell$ facilitators and  $N=100, b=1, c=0.8, d=0.2$.
Results 
 of stochastic simulations ($\circ$) for   $\phi^{\textsf{C}}$, 
analytical prediction (\ref{phi1C}) (solid) and $[N(1-z)]^{-1}$ (dashed), while $\times$'s show
$-\kappa\delta f(1)>0$   
when $z>{\widetilde z}=0.535$ [$\kappa=1$ (top) and $\kappa=4$ (bottom)]. 
Top: For strong selection intensity $s=1$, 
selection favors the invasion and replacement of ${\textsf{D}}$ by ${\textsf{C}}$
when $z>z^*=0.68$, see text.  
Bottom: For  $s=0.1$, one also finds $z^*=0.68$. 
Inset: $z^*$ vs. $Ns$ with $s=10^{-4} - 1$, $N=100$ ($\triangle$)
and  $N=1000$ ($\diamond$); we find $z^*\approx 0.684\pm 0.016$.
}
 \label{Fig1}
\end{figure}
In finite but large populations, when $N(b-c)(c-d)\gg 1$, the replacement condition (\ref{replacementPD})
is therefore satisfied  when the density $z$ of facilitators exceeds the critical value $z^*=\ell^*/N$,
with $\widetilde{z} \leq z^* \leq r$. Furthermore,  with (\ref{phiCphiDPD}) one has
$\phi^{\textsf{C}}>\phi^{\textsf{D}}$ in large populations  
 when  $x^*>(1-z)/2$  yielding $z>(c+d)/(2b-c+d)$.

One thus  distinguishes the following regimes:
  \begin{itemize}
   \item When  $z< {\widetilde z}$: Defection is   an ESS$_{N}^{\ell}$.
   \item When  $z> {\widetilde z}$: Defection is no longer an ESS$_{N}^{\ell}$.
   \item When ${\widetilde z}<z \leq z^*$: Selection favors the invasion (not the replacement) of defection 
   by cooperation. In this regime a single cooperator is more likely to go extinct than in the absence of selection.
    When $N(b-c)(c-d)\gg 1$ and $s>0$, one has ${\widetilde z}\leq z^*\leq r$ ($r=c/b$ is the cost-to-benefit ratio).
   \item When  $z> z^*$: Selection favors both the invasion and replacement  of defection by cooperation.
  \item When $z>(c+d)/(2b-c+d)$ and $N(2b-c+d)\gg 1$: ${\textsf{C}}$ is more likely to replace ${\textsf{D}}$ than vice versa.
  \end{itemize}
In PD games with $c>d$ and $z>0$, {\it the replacement condition 
(\ref{replacementPD}) is more stringent than (\ref{invasionPD}) and dictates when
selection favors the invasion and replacement of ${\textsf{D}}$ by ${\textsf{C}}$.}
 It is worth noting that the critical value 
  $z^*$ cannot be inferred from the rate equation (\ref{MF}) or from the payoffs (\ref{payoffs_PD}), but 
is determined by the nontrivial solution of (\ref{replacementPD}).

  These findings  are illustrated in Fig.~\ref{Fig1} for strong selection strength $s=1$ and for $s=0.1$. 
  In Fig.~\ref{Fig1},  the outcomes of stochastic 
simulations (obtained
with the Gillespie algorithm~\cite{Gillespie}) perfectly agree with the theoretical expression (\ref{phi1C}), with
 $\phi^{\textsf{C}}>[N(1-z)]^{-1}$ for $z \geq z^*$ and $z^*\leq r$, as predicted by (\ref{replacementPD}).
 The inset of Fig.~\ref{Fig1} illustrates  that the  critical
 value $z^*\approx 0.684\pm 0.016$ is obtained when $Ns=0.01-1000$, which indicates a very weak influence
of the  selection strength  on the value of  $z^*$.

 \subsection*{4.2. The case of positive payoff for mutual defection $(c< d)$} 
 When the payoff for mutual defection is positive $d-c> 0$, one has $\alpha< 0$
 and  $\psi_{\ell}(n)<0$ when $z>{\widetilde z}$. The replacement condition (\ref{replacementPD}) is 
 therefore always satisfied when  $\delta f(1)<0$ and (\ref{invasionPD}) is fulfilled.
  This implies that (\ref{invasion}) is more stringent than (\ref{replacementPD}) and
  {\it selection favors the invasion and replacement of $\textsf{D}$ by $\textsf{C}$
 when  $z>\widetilde{z}=[d+(b+c-2d)N^{-1}]/(b+d-c)$.} Furthermore, proceeding as in Sec.~4.1, one finds that
 the replacement condition (\ref{replacementPD})
is satisfied when $z> z^*$, where $0\leq z^*\leq r<\widetilde{z}$.
Moreover,  with (\ref{phiCphiD})  one still finds that 
$\phi^{\textsf{C}}>\phi^{\textsf{D}}$ in large populations  when  $z>(c+d)/(2b+d-c)$.

\begin{figure}
\includegraphics[width=3.6in, height=2.4in,clip=]{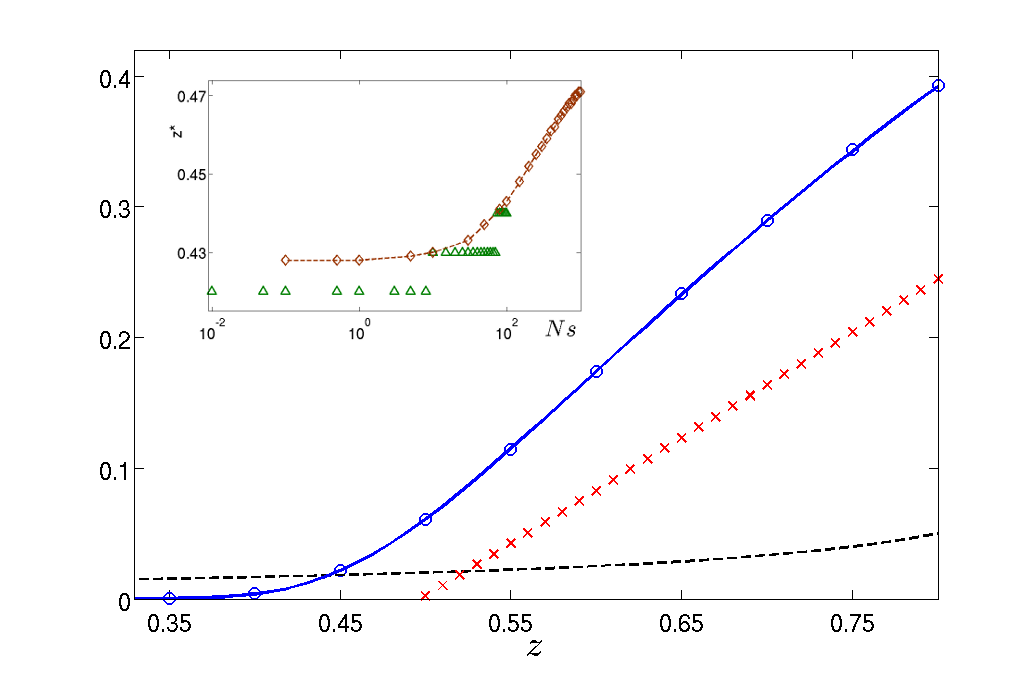}
\includegraphics[width=3.6in, height=2.4in,clip=]{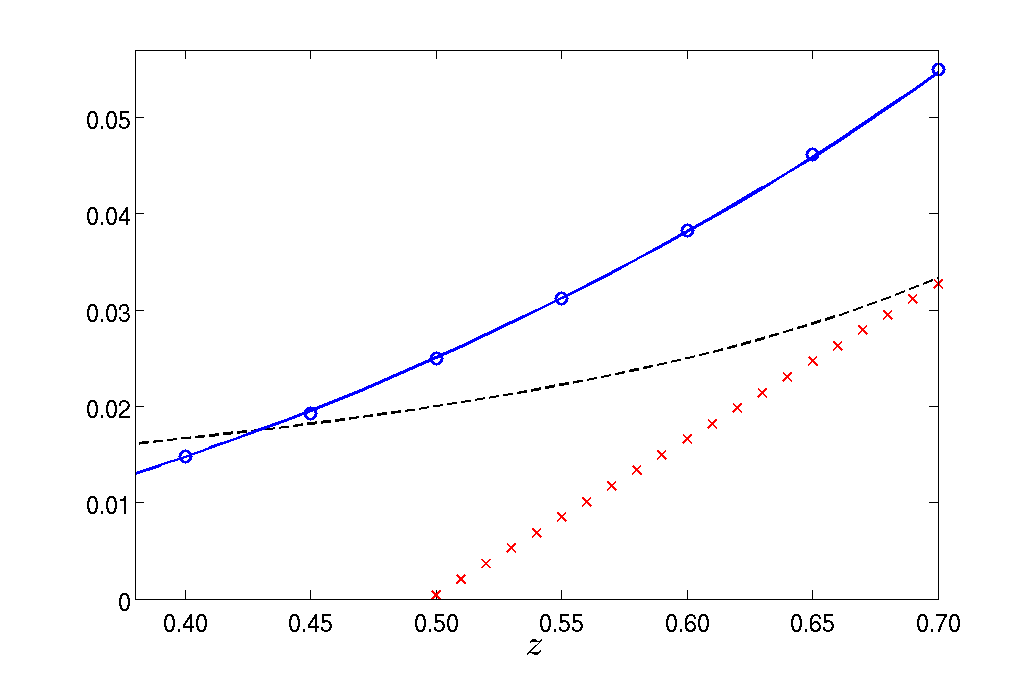}
\caption{{\it (Color online)}. 
As in Fig.~\ref{Fig1} with parameters $N=100, b=1, c=0.2, d=0.8$.
Results 
 of stochastic simulations ($\circ$) for   $\phi^{\textsf{C}}$, 
analytical prediction (\ref{phi1C}) (solid) and $[N(1-z)]^{-1}$ (dashed), while $\times$'s
 show
$-\kappa\delta f(1)>0$ when  $z>{\widetilde z} =0.4975$
[$\kappa=0.5$ (top) and $\kappa=1$ (bottom)]. Top: For strong selection intensity $s=1$, 
selection favors the replacement of ${\textsf{D}}$ by ${\textsf{C}}$
when $z>z^*=0.44$, see text.  
Bottom: For  selection intensity $s=0.1$, one has $z^*=0.43$. 
Inset: $z^*$ vs. $Ns$ with $s=10^{-4} - 1$, $N=100$ ($\triangle$)
and  $N=1000$ ($\diamond$); we find $z^*\approx 0.42-0.47$.
}
 \label{Fig2}
\end{figure}

 The invasion and replacement conditions (\ref{invasionPD}) and (\ref{replacementPD}) are 
 illustrated in Figure~\ref{Fig2}, where the theoretical prediction (\ref{phi1C}) perfectly agrees with 
 the results of stochastic simulations, and can be summarized as follows:
  \begin{itemize}
   \item When $z \leq z^*$: Defection is   an ESS$_{N}^{\ell}$.
   \item When $z>  z^*$: Defection   is no longer an ESS$_{N}^{\ell}$
and  $r\leq z^*<\widetilde{z}$.
    \item When $z^*< z<{\widetilde z}$: Selection favors the replacement (not the invasion) of defection 
   by cooperation. In this regime a single cooperator is more likely to fixate than in the absence of selection.
   \item When $z> {\widetilde z}$: Selection favors both the invasion and replacement  of defection by cooperation.
    \item  When $z>(c+d)/(2b+d-c)$ and $N(2b+d-c)\gg 1$: ${\textsf{C}}$ is more likely to replace ${\textsf{D}}$ than vice versa.
   \end{itemize}

 The inset of Fig.~\ref{Fig2} indicates again that the selection intensity has
a weak  influence on the value of $z^*$: in Fig.~\ref{Fig2}, one finds
 $z^*\approx 0.42 - 0.47$ when $Ns=0.01-1000$.
  
 It is  worth noting that  in the simple and special case of ``equals-gains-from-switching'' PD games, with $c=d$, 
considered in Ref.~\cite{MM12}, one has ${\widetilde z}=z^*$~\cite{ackn} and in this case  
${\textsf{C}}$ is more likely to invade and
replace ${\textsf{D}}$ when $z>r$ and $N\gg 1$, see (\ref{phiCphiDPD}),
i.e. when the density of facilitators exceeds the cost-to-benefit ratio.

 \vspace{0.3cm}
 
 We have thus shown that in PD games with $\ell$ cooperator facilitators 
selection favors the invasion and replacement of defection by cooperation
when the fraction of cooperation facilitators  exceeds 
a critical value $z^*$ (with $z>z^*>\widetilde{z}$)
imposed by the replacement condition (\ref{replacementPD}) $c> d$.
When $c\leq d$, the invasion condition 
 (\ref{invasion}) dictates that 
 $z>\widetilde{z}$ (with $\widetilde{z}\geq z^*$)
for both the  invasion and replacement of defection to be favored.

\section*{5. When does selection favor cooperation in snowdrift games with facilitators?}
 The expected payoffs of SD games with $\ell$ cooperator facilitators are given by (\ref{payoffs_SD})
 with $b>c+d>0$. One thus has
 $\delta f(j)=s[\alpha (j/N)+ \beta_{\ell}]$, with 
 $\alpha=(c+d)N/(N-1)$ and $\beta_{\ell}=[b-dN-(b-c-d)\ell]/(N-1)$. 

 It is again instructive to start by considering the mean field dynamics  (\ref{MF})  
 characterized by two absorbing fixed points ($x=0$ and $x=1-z$), as well as a coexistence fixed point 
 $x^*=[d+(b-c-d)z]/(c+d)$ when $z<c/b$. At mean field level, two regimes have to be considered: 
\begin{itemize}
 \item When $z<c/b$, the interior fixed point $x^*$ is the only attractor of (\ref{MF}) that predicts the
 long-lived coexistence of cooperators and defectors.
 \item When $z\geq c/b$, the absorbing state $x=1-z$ is the only attractor of (\ref{MF}). This
corresponds to the situation where cooperation prevails over defection.
\end{itemize}

This analysis indicates that  when the number of facilitators is raised from $0$ to $c/b$, the interior fixed point $x^*$ is an 
attractor and an initial small density of cooperator increases to reach the value $x^*$. 
Furthermore, when the fraction of facilitators $z\geq c/b$, any initial number of cooperators grows and
quickly replace all the defector. At mean field level, it is thus clear that a fraction $z\geq c/b$ of facilitators
leads to a sustained level of cooperation in SD games. This picture is significantly 
altered by the demographic fluctuations arising in finite populations, where  $x^*$ is  metastable~\cite{snowdrift1,snowdrift2,snowdrift3} and
 a careful analysis has to be carried out to determine under which circumstances the selection 
 favors cooperation replacing defection.

Since the mean 
 field analysis has revealed that in the absence of fluctuations there is a sustained level of cooperation for $z\geq 0$, 
 we need to focus  on the replacement condition (\ref{replacement}) to understand the combined influence of selection and 
demographic noise. In fact, the invasion condition  (\ref{invasion}) here  yields $z>{\widetilde z}=
-d^{-1}+[(b+c+d)/(b-c-d)]/N$ which simply reduces to $z\geq 0$ and is always satisfied when $N(b-c-d)\gg 1$.

For SD games in the presence of $\ell$ facilitators, the condition under which cooperation is favored by
selection is thus given by (\ref{replacementSD}) which here reads
 \begin{eqnarray}
  \label{replacementSD}
  N-\ell-1&>&\sum_{n=1}^{N-\ell-1}e^{n \psi_{\ell}(n)}, \quad \text{with}\\
  \label{replacementSD-2}
  \psi_{\ell}(n)&=&\delta f(1)+s\frac{(c+d)(n-1)}{2(N-1)}\nonumber\\
  &=&s \left[\frac{(c+d)(n+1)}{2(N-1)}+\beta_{\ell}\right]. \nonumber
 \end{eqnarray}

One can 
 analyze (\ref{replacementSD}) as in Sec.~4 and distinguish
between the cases $c\geq d$ and $c<d$. In fact,  from (\ref{phiCphiD}) we infer
that in large populations $\phi^\textsf{C}>\phi^\textsf{D}$ when $x^*>(1-z)/2$, which
yields $z>(c-d)/(2b-c-d)$ when $c\geq d$, whereas $\textsf{C}$ is always more likely to replace 
$\textsf{D}$ than vice versa when  $c< d$.

\begin{figure}
\includegraphics[width=3.6in, height=2.4in,clip=]{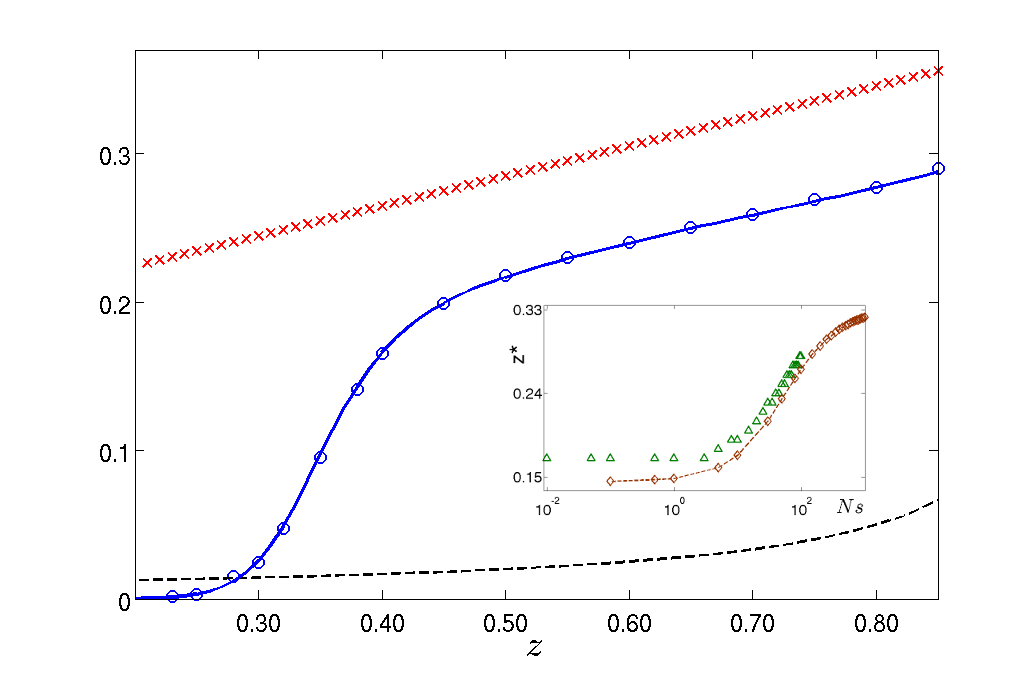}
\includegraphics[width=3.6in, height=2.4in,clip=]{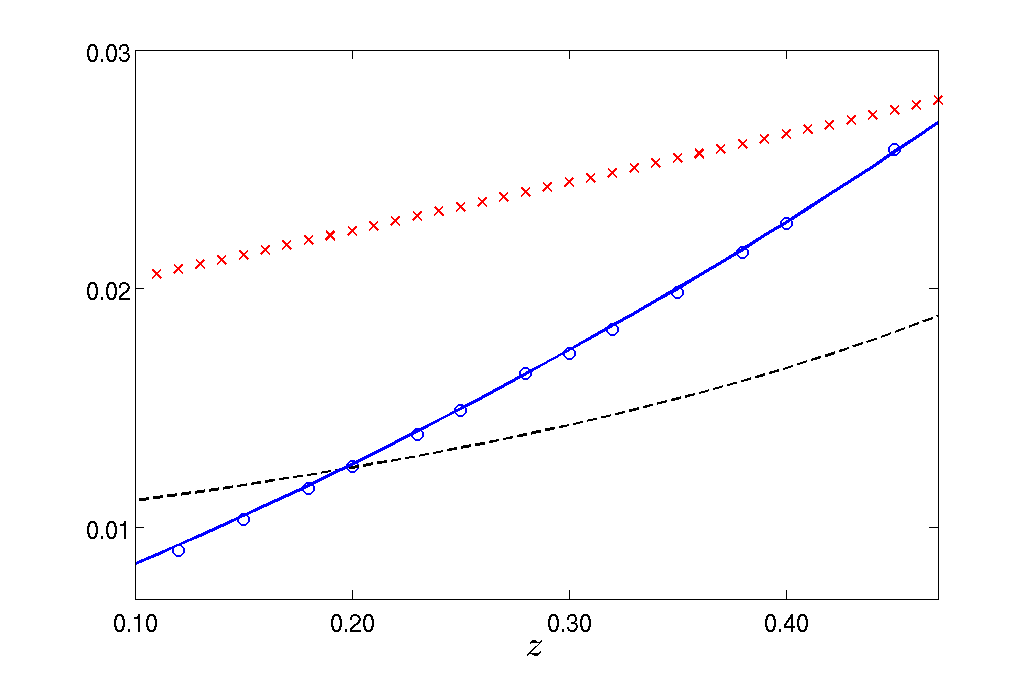}
\caption{{\it (Color online)}.
$\phi^{\textsf{C}}$ and $-\delta f(1)$ vs. $z=\ell/N$
in SD games with  
$\ell$ facilitators and  $N=100, b=1, c=0.6, d=0.2$. Results 
 of stochastic simulations ($\circ$) for   $\phi^{\textsf{C}}$, 
analytical prediction (\ref{phi1C}) (solid) and $[N(1-z)]^{-1}$ (dashed), while $\times$'s show
$-\delta f(1)>0$ for all values of $z\geq 0$. Top: For  $s=1$, 
selection favors  ${\textsf{C}}$ replacing ${\textsf{D}}$ 
when $z>z^*=0.28$, see text.  
Bottom: For $s=0.1$, we find $z^*=0.19$. 
Inset: $z^*$ vs. $Ns$ with $s=10^{-4} - 1$, $N=100$ ($\triangle$)
and  $N=1000$ ($\diamond$).}
 \label{Fig3}
\end{figure}

The more interesting situation arises when $c\geq d$ and  $N(b-c)\gg 1$. In this case, {\it  there is a critical 
fraction of facilitators $z^*$, with  $0\leq z^*\leq c/b$, above which (\ref{replacementSD}) is satisfied 
and selection favors $\textsf{C}$ invading and replacing $\textsf{D}$}. 
 The  replacement condition (\ref{replacementSD}) and $\delta f(1)<0$ for $c\geq d$ are illustrated in Fig.~\ref{Fig3}, in which
 an excellent agreement is reported between (\ref{phi1C}) and  stochastic simulations: it is 
found that $\phi^\textsf{C}>[N(1-z)]^{-1}$ when $z>z^*\geq c/b$ as predicted by (\ref{replacementSD}).
The inset of Fig.~\ref{Fig3} illustrates that the critical value 
$z^*$ grows with the selection intensity $s$ and population size $N$
as a monotonic  function of the scaling variable $Ns$.

When $c<d$ and $z<c/b$, the coexistence fixed point 
is closer to the absorbing state $x=1-z$ than to $x=0$. In this case, 
when $N(d-c)\gg 1$, {\it  the replacement condition (\ref{replacementSD}) is satisfied for $z\geq 0$, i.e. 
 even in the absence of facilitators, and selection always favors the invasion and replacement of $\textsf{D}$ by 
$\textsf{C}$}.

 In summary, SD games with $\ell$ cooperation facilitators 
are characterized by the following regimes:
  \begin{itemize}
   \item When $z\geq 0$ and $(b-c-d)N\gg1$: Defection is never an ESS$_{N}^{\ell}$ since $\delta f(1)<0$ and the invasion of 
${\textsf{D}}$ by ${\textsf{C}}$ is favored.
   \item When  $c\geq d$ and $0\leq z \leq z^*$:  Selection favors the invasion (not the replacement) of defection 
   by cooperation. In this regime, the fixation probability of a single cooperator is
   lower than in the absence of selection ($s=0$). 
   \item When $c\geq d$ and  $z >  z^*$, with $N(c-d)\gg 1$: Selection favors the invasion and replacement  of defection by cooperation.
   The critical value  $z^*\leq c/b$ is an increasing function of $Ns$ (see inset of Fig.~\ref{Fig3}).
   \item When $c\geq d$ and $z>(c-d)/(2b-c-d)$: ${\textsf{C}}$ is  more likely to replace ${\textsf{D}}$ 
   than vice versa in large populations.
   \item  When $c< d$ and $N(c-d)\gg 1$: Selection favors the invasion and replacement  of defection by cooperation 
even in the absence of facilitators and ${\textsf{C}}$ is always more likely to replace ${\textsf{D}}$ than vice versa, i.e. for any $z\geq 0$. 
   \end{itemize}

\section*{6. When does selection favor cooperation in stag-hunt games with facilitators?}
 The expected payoffs of SH games (\ref{PM_SH}) with $\ell$ cooperator facilitators are given by (\ref{payoffs_SH}), and the 
 fitness difference reads $\delta f(j)=s[\alpha (j/N)+ \beta_{\ell}]$, with 
 $\alpha=-(c+d)N/(N-1)$ and $\beta_{\ell}=[dN-(b+d)\ell+ b+c-d]/(N-1)$. 

 Useful information is provided by  the study of the mean field rate equation (\ref{MF}) 
 characterized by an unstable interior fixed point $x^*=[d-(b+d)z]/(c+d)$ separating the 
 absorbing states $x=0$ and $x=1-z$ when  $0\leq z<d/(b+d)$. There are two 
 mean field regimes: 
\begin{itemize}
 \item When $0\leq z<d/(b+d)$,  $x^*$ exists and is unstable, whereas $x=0$ (no 
  $\textsf{C}$'s) and $x=1-z$ (no   $\textsf{D}$'s) are both stable (bistability).
 Since $x^*<(1-z)/2$, the state $x=1-z$ has a larger basin of attraction than $x=0$
and $\textsf{C}$ is therefore risk-dominant:
 (\ref{MF}) leads to a state with no 
  $\textsf{D}$'s when $x(0)>x^*$
  ~\cite{EGT4,EGT5,EGT6,staghunt1,staghunt2,weaksel1,weaksel2}. 
 \item When $z \geq d/(b+d)$, the absorbing state $x=1-z$ is the only attractor of (\ref{MF}) whereas
 $x=0$ is unstable and there is no interior fixed point. This situation corresponds to 
$\textsf{C}$ prevailing over $\textsf{D}$.
\end{itemize}

The mean field analysis reveals that the absorbing states are stable in SH games
and separated by the unstable coexistence fixed point $x^*$ when 
the fraction of facilitators is raised from $0$ to $z=d/(b+d)$. 
In this situation, a small number of cooperators would 
quickly die out under the mean field dynamics (\ref{MF}) and it is necessary that $z\geq d/(b+d)$
for cooperation to be able to spread in infinitely large populations.

\begin{figure}
\includegraphics[width=3.6in, height=2.4in,clip=]{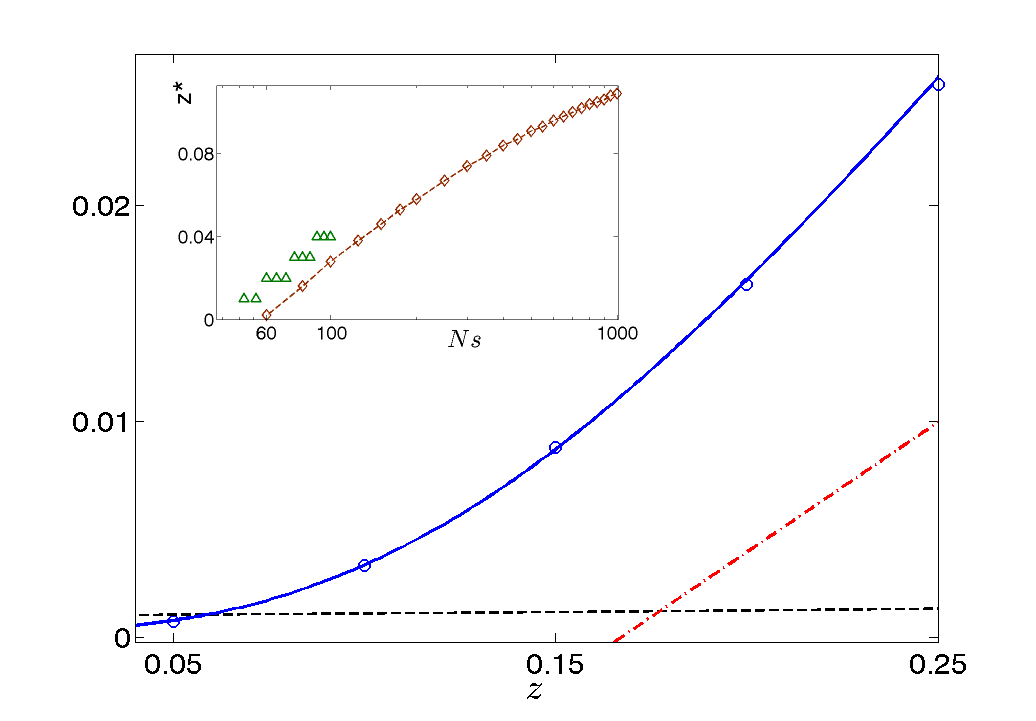}
\includegraphics[width=3.6in, height=2.4in,clip=]{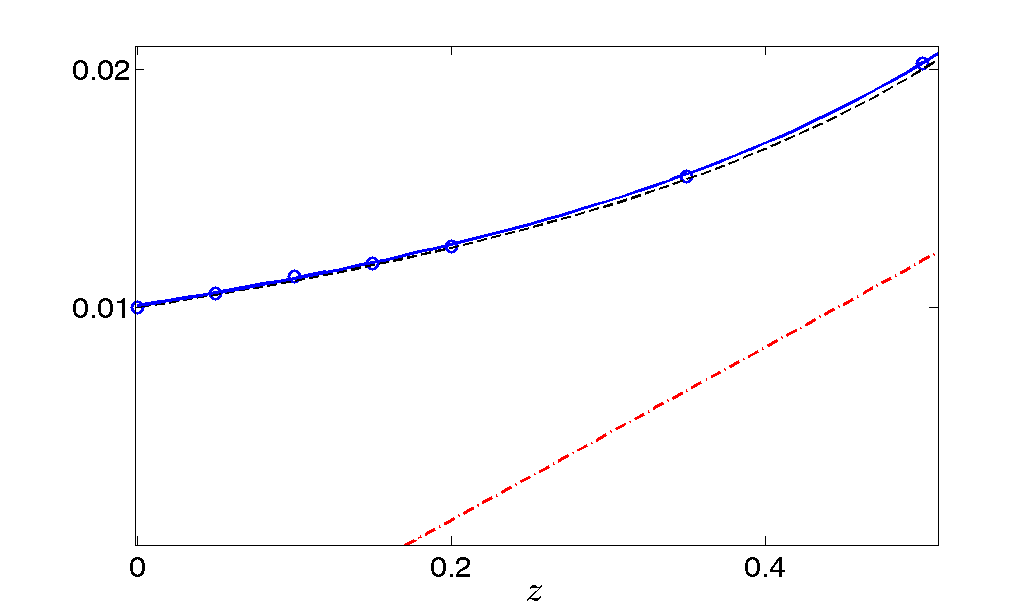}
\caption{{\it (Color online)}.
$\phi^{\textsf{C}}$ and $-\delta f(1)>0$ vs. $z=\ell/N$
in SH games with  
$\ell$ facilitators and  $b=1, c=0.8, d=0.2$.
Here, $x^*=0.2-1.2z$.
Results 
 of stochastic simulations ($\circ$) for   $\phi^{\textsf{C}}$, 
analytical prediction (\ref{phi1C}) (solid) and $[N(1-z)]^{-1}$ (dashed); 
while the dashed-dotted lines 
 show
$-\kappa\delta f(1)>0$ when  $z>{\widetilde z}$
[$\kappa=1$ (top) and $\kappa=30$ (bottom)].
Top: When $s=0.2$ and $N=1000$, ${\widetilde z}\approx 0.167$, and 
selection favors ${\textsf{C}}$ replacing ${\textsf{D}}$ 
when $z>z^*\approx 0.059$.   
Bottom: When $s=0.1$ and $N=100$, selection always favors the replacement of defection and
${\widetilde z}\approx 0.172$.
Inset: 
 $z^*$ vs. $Ns$ with $s=10^{-4} - 1$, $N=100$ ($\triangle$)
and  $N=1000$ ($\diamond$); when $Ns\lesssim 60$, 
the replacement condition (\ref{replacementSH}) is satisfied even when $\ell=0$, see text.
}
 \label{Fig4}
\end{figure}

In the absence of facilitators ($\ell=0$),  SH games  are always characterized by the bistability of 
$\textsf{C}$ and $\textsf{D}$ (two strict NEs) and, under weak selection and in populations of large but finite size
($Ns\ll 1$ and $N\gg 1$), it has been shown that selection  favors the the replacement of  $\textsf{D}$ by $\textsf{C}$
when  $x^*<1/3$, which here  means $c>2d$, i.e. when the basin of attraction of $\textsf{C}$ is more than twice that of $\textsf{D}$
 (the so-called ``1/3-rule'')~\cite{ESSN,EGT4,EGT6,staghunt1,weaksel1}.

In SH games with cooperation facilitators, the invasion 
 condition $\delta f(1)<0$ (\ref{invasion}) leads to $z> \widetilde{z}$, where 
 \begin{eqnarray}
  \label{invasionSH}
 \widetilde{z}= \frac{d}{b+d}+\left(\frac{b-2d}{b+d}\right)\frac{1}{N},
 \end{eqnarray}
which coincides with the mean field requirement $z>d/(b+d)$ when $N\to \infty$. Moreover, 
the replacement condition (\ref{replacement}) yields
 \begin{eqnarray}
  \label{replacementSH}
   N-\ell-1&>&\sum_{n=1}^{N-\ell-1}e^{n \psi_{\ell}(n)}, \quad \text{with} \\
  \label{replacementSH-2}
  \psi_{\ell}(n)&=&\delta f(1)-s\frac{(c+d)(n-1)}{2(N-1)}\nonumber\\
  &=&s \left[-\frac{(c+d)(n+1)}{2(N-1)}+\beta_{\ell}\right].\nonumber
 \end{eqnarray}
Since $z>{\widetilde z}$ guarantees $\delta f(1)<0$ 
and thus  $\psi_{\ell}(n)<0$, it is clear that (\ref{replacementSH}) is satisfied when
the invasion condition (\ref{invasion}) is fulfilled: {\it $z>{\widetilde z}$   is therefore more stringent than 
(\ref{replacementSH}) and dictates that  selection favors the
invasion and replacement of defection by cooperation when the density of facilitators in the population
exceeds (\ref{invasionSH})}. As a consequence, the replacement condition (\ref{replacementSH})
is satisfied when $z>z^*$, with $z^*\leq {\widetilde z}$.
Furthermore, with (\ref{phiCphiD}), (\ref{alpha}) and (\ref{beta}), we find that 
$\phi^{\textsf{C}}>\phi^{\textsf{D}}$ when $x^*<(1-z)/2 -[(b+c+d)/(c+d)]/(2N)$. In 
large populations ($N(c+d)\gg 1$) this yields $x^*<(1-z)/2$, a condition which is here always satisfied, 
and therefore in SH games (\ref{PM_SH}) $\textsf{C}$ is always risk-dominant and more likely to replace $\textsf{D}$ 
than vice versa.

Guided by the above mean field analysis and discussion of the 1/3-rule, in our analysis
we distinguish the case where the selection strength is finite, with $N\gg 1$ and $Ns\gg 1$,
from the situation where the population size is large (but finite) and the selection intensity is weak 
($N\gg 1$ and $Ns\ll 1$).

\begin{figure}
\includegraphics[width=3.6in, height=2.4in,clip=]{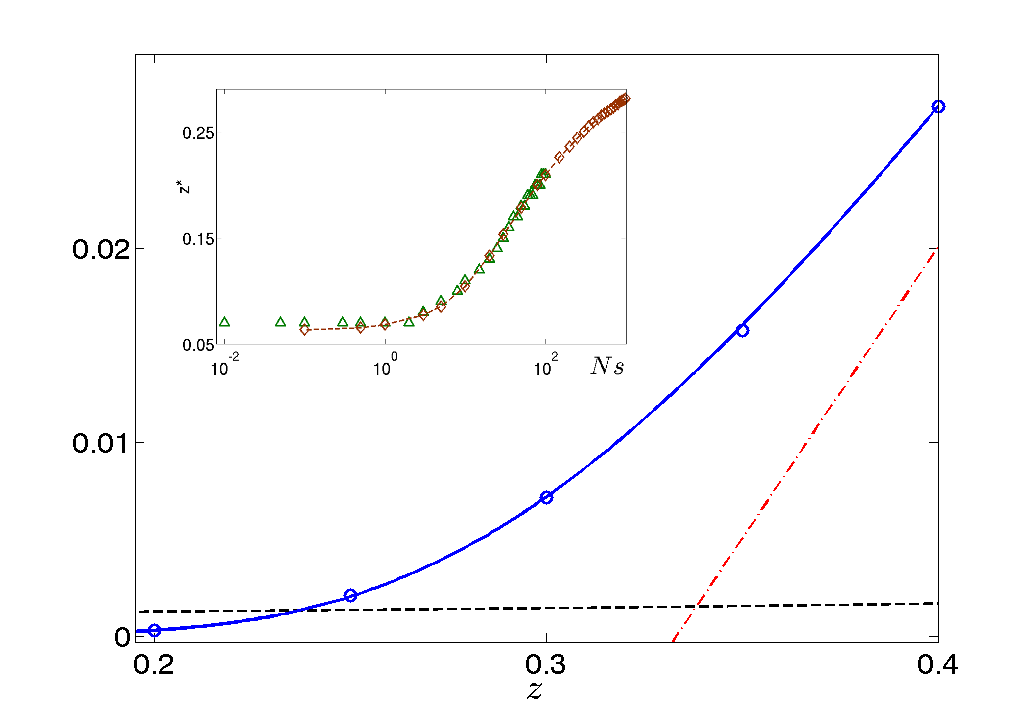}
\includegraphics[width=3.6in, height=2.4in,clip=]{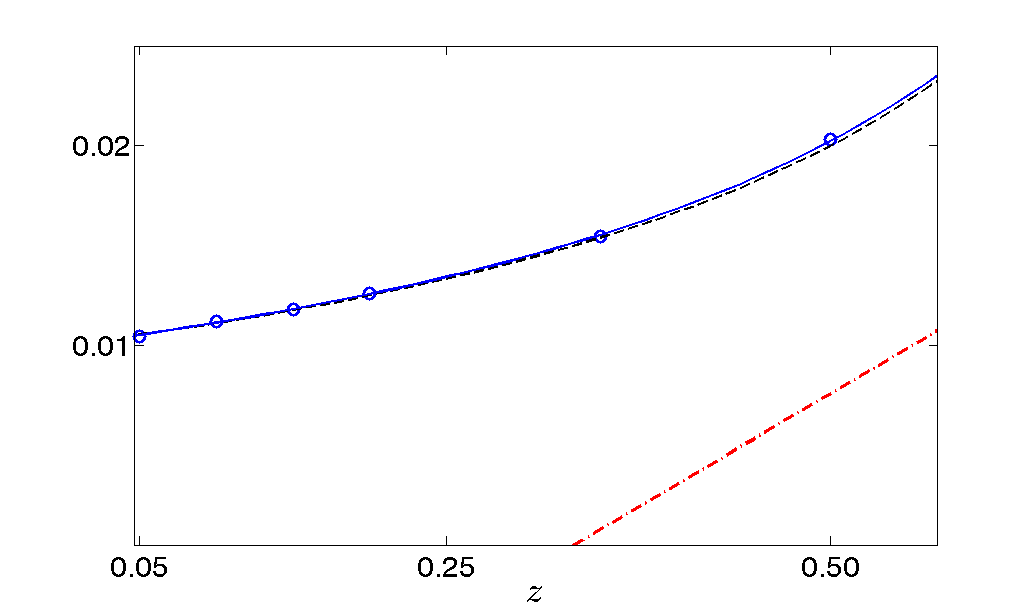}
\caption{{\it (Color online)}. As in Fig.~\ref{Fig4} with parameters $b=1, c=0.8, d=0.5$.
Here, $x^*=(5/13)(1-3z)$.
Results 
 of stochastic simulations ($\circ$) for   $\phi^{\textsf{C}}$, 
analytical prediction (\ref{phi1C}) (solid) and $[N(1-z)]^{-1}$ (dashed); 
while the dashed-dotted lines 
 show
$-\kappa\delta f(1)>0$ when  $z>{\widetilde z}$
[$\kappa=1$ (top) and $\kappa=30$ (bottom)].
Top: When $s=0.2$ and $N=1000$,   ${\widetilde z}\approx 0.33$, and
selection favors ${\textsf{C}}$ replacing ${\textsf{D}}$ 
when
$z>z^*\approx 0.24$. 
Bottom: When $s=0.001$ and $N=100$, one has $z^*\approx 0.06$ and ${\widetilde z}\approx 0.33$, see text.
Inset: 
 $z^*$ vs. $Ns$ with $s=10^{-4} - 1$, $N=100$ ($\triangle$)
and  $N=1000$ ($\diamond$).}
 \label{Fig5}
\end{figure}

When $N\gg 1$ and $Ns\gg 1$, there is a critical density of facilitators $z^*$, with $0<z^*\leq {\widetilde z}$
above which (\ref{replacementSH}) is satisfied: {\it selection favors the replacement of defection by cooperation
when $z>z^*$.}   
This is illustrated in Figs.~\ref{Fig4} and \ref{Fig5} (top)
where the results of stochastic simulations for $\phi^\textsf{C}$
coincide with the prediction (\ref{phi1C}) and it is
confirmed that (\ref{replacementSH}) is satisfied when  $z>z^*$, with $0<z^*\leq {\widetilde z}$,
while  both (\ref{invasion}) and 
(\ref{replacementSH}) are  satisfied when 
$z> {\widetilde z}$. The critical value $z^*$ grows  with $s$
and admits ${\widetilde z}$ as an upper-bound. In the insets  of Figs.~\ref{Fig4} and ~\ref{Fig5}
 the dependence of $z^*$ on  $s$ and $N$ is shown to be given by monotonically growing 
 functions of the scaling variable $Ns$.

When the coexistence fixed point $x^*$ is close to the absorbing state $x=0$ 
and $Ns\ll 1$ one finds that selection favors the replacement of defection 
even in the absence of facilitators, as illustrated in 
Fig.~\ref{Fig4} (bottom). This can be rationalized by generalizing the 
1/3-rule in the presence of facilitators~\cite{ESSN,EGT4,EGT6,staghunt1,weaksel1}.
In fact, for large populations and weak selection, when
 $N(1-z)\gg 1$ and $Ns\ll 1$, one can write
$\sum_{n=1}^{N-\ell-1}e^{n \psi_{\ell}(n)}=N(1-z)\left[1+(Ns/6)(1-z)
\left\{2d-c-z(3b-c+2d)+{\cal O}(Ns)\right\}\right]$. The replacement condition (\ref{replacementSH})
thus becomes
\begin{eqnarray}
\label{weak}
z>\frac{2d-c}{3b-c +2d}.
\end{eqnarray}
Hence, {\it when $c>2d$ the replacement of defectors is favored by (weak) selection 
for any $z\geq 0$, i.e. even in the absence of facilitators, while 
$z^*=(2d-c)/(3b-c+2d)$ when $c<2d$.} The result (\ref{weak})
can also be rewritten as the following condition on the location of the unstable coexistence fixed point:
\begin{eqnarray}
 \label{third}
 x^*<1/3 - (z/9)[1+(c/d)],
\end{eqnarray}
where  terms of order ${\cal O}(z^2)$ have been neglected and the 1/3-rule is recovered 
when $z=0$. We infer from the inset of Fig.~\ref{Fig4}, that when 
$Ns\gtrsim 60$ a density $z>z^*$
of facilitators is necessary to 
 favor the replacement of defection.
In Fig.~\ref{Fig5}~(bottom), selection is found to (slightly) favor the replacement of defectors by cooperators
when $z>z^*\approx 0.06-0.07$ in good agreement with (\ref{weak},\ref{third}).

In summary, selection favors the invasion and replacement of defection by cooperation in stag-hunt games in the presence 
of $\ell$ facilitators when the invasion condition  $z> {\widetilde z}$, with (\ref{invasionSH}),  is satisfied, 
and one distinguishes the following scenarios:
  \begin{itemize}
   \item When$Ns\gg 1$ and  $z\leq z^*$: Defection is an ESS$_{N}^{\ell}$.
   \item When $Ns\gg 1$ and  $z>  z^*$: Defection is no longer an ESS$_{N}^{\ell}$. 
      The critical density  of 
   facilitators $z^*$ is an increasing function of $Ns$, with
$0<z^*\leq {\widetilde z}$. 
   \item When $z^*<z\leq {\widetilde z}$: Selection favors the replacement (not the invasion)
   of defection by cooperation. In this regime, the fixation probability of a single cooperator is
   higher than in the absence of selection ($s=0$). When $Ns\ll 1$ (weak selection) and $N(1-z)\gg 
   1$,  $z^*=(2d-c)/(3b-c+2d)$ if $c<2d$, while  selection favors the replacement of defectors even when 
   $z=0$ if $c>2d$.
   \item When $ z> {\widetilde z}$:  Selection favors both the invasion and replacement  of
 defection by cooperation.
    \item  When $N(c+d)\gg 1$ and $z\geq 0$
 $\textsf{C}$ is always risk-dominant in SH games  (\ref{PM_SH}) with cooperation facilitators, and  more likely to replace $\textsf{D}$ than vice versa 
    ($\phi^\textsf{C}>\phi^\textsf{D}$).
   \end{itemize}

\section*{7. Conclusion}
We have studied the combined influence of random fluctuations
and selection pressure on the evolutionary dynamics of three classes of two-strategy games
in the presence of cooperation facilitators. More precisely, we have considered the evolution
of the prisoner's dilemma (PD), snowdrift (SD) and stag-hunt (SH) games in a
well-mixed population of size $N$ comprising
$\ell$ cooperation facilitators. These are individuals enhancing the cooperators' reproductive potential
(relative to defectors' fitness). %
By computing the exact cooperation fixation probability and by
analyzing the mean field dynamics, we have investigated the circumstances 
under which selection favors the invasion and replacement of defection by cooperation.
Invasion is favored when cooperators have a higher fitness
than defectors, whereas selection favors the replacement of defectors
when the fixation probability of a single cooperator is higher than  $(N-\ell)^{-1}$ obtained under neutral dynamics.
We have also determined the conditions under which cooperation is more likely to replace defection than vice versa.
Interestingly, the invasion and replacement conditions lead to various scenarios when the fraction of facilitators 
present in the population varies. 
In PD and SD games, selection favors the replacement of
defection by cooperation  when the density of facilitators exceeds a critical value $z^*$
that depends on the model's parameters and  admits the cost-to-benefit ratio as an upper-bound.
Selection also favors cooperators  invading defectors in PD games with negative payoff for mutual defection when the 
density of facilitators exceeds $z^*$.
When the  payoff for mutual defection is positive in PD games, selection favors both the replacement and 
invasion of cooperation by defection when cooperators have higher  fitness than defectors.
In SD games,  defection is generally not  evolutionary stable in large populations and selection favors its 
invasion and replacement by cooperation when the density of facilitators is greater than $z^*$.
In large populations, the critical value $z^*$ is found to display a weak dependence on the population size 
$N$  and selection strength $s$ in PD games, whereas in SD and SH games $z^*$ grows nontrivially with the scaling variable $Ns$.
In SH games, the invasion and replacement of defection is favored whenever cooperators have a higher fitness than 
defectors. 
We have also generalized the so-called 1/3-rule and shown that in SH games  the replacement 
of defection under weak selection occurs when the unstable mean field coexistence fixed point is below a critical threshold
that depends on the density of facilitators. %
\\
These findings show that the  presence of  cooperation facilitators can drastically alter 
the dynamics of paradigmatic models of social dilemmas via various scenarios, and can possibly be seen as an alternative mechanism
to try and explain examples of cooperative behavior. It would be interesting to investigate 
the influence of cooperation facilitators on the evolution of spatially structured populations on complex 
networks, a subject that has recently received much attention, see e.g.~\cite{EGT5,nets1,nets2,nets3}.


\begin{thebibliography}{10}
\bibitem{ValInt}
Since  
 $\psi_{\ell}(n)$ given by (\ref{replacementPD-2}) increases with  $n$ and decreases with $\ell$, 
 while $\psi_{\lfloor \widetilde{\ell}\rfloor}(n)\geq 0$ 
and  $\psi_{\lceil Nr\rceil}(n)\leq 0$ when  
$N(b-c)(c-d)\gg 1$, one obtains $\sigma_{\rm PD}(N,\lfloor \widetilde{\ell}\rfloor)>0$ 
and  $\sigma_{\rm PD}(N,\lceil Nr\rceil)< 0$. One therefore generally 
has  $\sigma_{\rm PD}(N,\ell)< 0$
for $\ell > \ell^*$, where $\lfloor \widetilde{\ell}\rfloor \leq \ell^*\leq \lceil Nr\rceil$,
as confirmed in Figs.~\ref{Fig1} and \ref{Fig2}.
\end{thebibliography}
\end{document}